\documentclass[aps,showpacs,amsmath,amssymb,nofootinbib,preprintnumbers]{revtex4}

\usepackage{graphicx}
\usepackage{color}

\def \be  {\begin{equation}}
\def \ee  {\end{equation}}
\def \bea {\begin{eqnarray}}
\def \eea {\end{eqnarray}}

\begin{document}

\preprint{ECTP-2011-01}

\title{The Hubble parameter in the early universe with viscous QCD matter and finite cosmological constant}
\author{A.~Tawfik}
\email{atawfik@cern.ch}
\affiliation{Egyptian Center for Theoretical Physics (ECTP), MTI University, Cairo-Egypt}

\begin{abstract}

The evolution of a flat, isotropic and homogeneous universe is studied. The background geometry in the early phases of the universe is conjectured to be filled with causal bulk viscous fluid and dark energy. The energy density relations obtained from the assumption of covariant conservation of energy-momentum tensor of the background matter in the early universe are used to derive the basic equation for the Hubble parameter $H$. The viscous properties described by ultra-relativistic equations of state and bulk viscosity taken from recent heavy-ion collisions and lattice QCD calculations have been utilized to give an approximate solution of the field equations. The cosmological constant is conjectured to be related to the energy density of the vacuum. In this treatment, there is a clear evidence for singularity at vanishing cosmic time $t$ indicating the dominant contribution from the dark energy. The time evolution of $H$ seems to last for much longer time than the ideal case, where both cosmological constant and viscosity coefficient are entirely vanishing. 

\end{abstract}

\pacs{95.30.Sf, 04.20.-q, 98.80.Jk, 04.20.-q}

\maketitle

\section{Introduction}
\label{sec:intro}

Dark energy and dissipative processes are supposed to play a very important role in the evolution of the early universe and therefore would have essential astrophysical and cosmological consequences~\cite{Tawfik:2008cd}. Much progress has been achieved in relativistic thermodynamics of dissipative fluids \cite{Ec40,LaLi87}. The widely used theoretical framework is the Israel and Stewart (IS) theory~\cite{Is76,IsSt76}, in which the causality is apparently conserved and the theory itself seems to be thermodynamically stable \cite{HiLi89,Ma95}. The dynamics and evolution of the early universe can be described by causal {\it bulk} viscous thermodynamics \cite{Tawfik:2010ht,Tawfik:2010mk,Tawfik:2009nh,Tawfik:2011bm}. But due to their complicated characteristics, very few exact solutions of the gravitational field equations are derived, especially in framework of full causal IS theory \cite{ChJa97,MaHa99a,MaHa99b,MaHa00a,MaTr97}. Also, it has been proposed that causal bulk viscous thermodynamics can model on a phenomenological level matter creation in the early universe \cite{ChJa97, MaHa99a}. Recently, the effects of bulk viscosity on the early universe with vanishing cosmological constant $\Lambda$ have been investigated \cite{Tawfik:2010ht,Tawfik:2010mk,Tawfik:2009nh,Tawfik:2011bm}. 

Basically, the mathematical difficulties of the non-linear and non-homogeneous differential equations result in limiting most investigations of dissipative causal cosmologies to homogeneity and isotropy Friedmann-Lemaitre-Robertson-Walker (FLRW) symmetry \cite{MaTr97}. The Einstein field equations for such models can be decoupled and therefore are reduced to an autonomous system of first order ordinary differential equations, which can be analyzed approximately \cite{CoHo95}. To have a reliable description, the background matter in FLRW model is assumed to be filled with dark energy and an ultra-relativistic viscous QCD matter. The bulk viscosity of the QCD matter and its equation of state have been deduced from recent heavy-ion collisions experiments and the lattice QCD simulations. In the present work, we study such a sophisticated system of non-linear and non-homogeneous differential equations at finite bulk viscosity and cosmological constant.

Recent RHIC results give a strong indication that hot dense matter is conjectured to be formed in the heavy-ion collisions experiments \cite{reff1}. Such an experimental evidence seems to agree with the {\it ''new state of matter''} as predicted in the lattice QCD simulations~\cite{reff5}. However, the experimentally observed elliptic flow in peripheral heavy-ion collisions would give an indication that a thermalized collective QCD matter has been produced. In a addition to that, the success of ideal fluid dynamics in explaining several experimental data, e.g. transverse momentum spectra of identified particles, the elliptic flow~\cite{reff6} shows that a {\it nearly} perfect fluid is likely to created and the quarks and gluons are likely to go through a relatively rapid equilibrium characterized with a thermalization time less than $1$~fm/c~\cite{mueller1}. According to recent lattice QCD simulations~\cite{mueller2}, the bulk viscosity $\xi$ is not negligible near the QCD critical temperature $T_c$. It has been shown that the bulk and shear viscosity at high temperature $T$ and weak coupling $\alpha_s$ are respectively given as $\xi\simeq\alpha_s^2 T^3/\ln \alpha_s^{-1}$ \cite{mueller3}. Such a behavior obviously reflects the fact that near $T_c$ QCD is far from being conformal. But at high $T$, QCD approaches conformal invariance, which can be indicated by low trace anomaly $(\rho-3p)/T^4$~\cite{karsh09}, where $\rho$ and $p$ are the energy and pressure density, respectively. 

The present work is devoted to investigate the effects of dark energy and bulk viscosity on the evolution of the Hubble parameter $H$ in the early universe, especially in the quark-gluon plasma (QGP) era, which is positioned within the range $0.2\leq T\leq10\;$GeV or $18.35\leq t \leq 0.0073\;$GeV$^{-1}$. This limitation is likely compatible with the recent lattice QCD simulations and the experimental evidences that QGP seems to remain strongly correlated, i.e. non-viscous, up to several $T_c$. Also, this sets on the upper and lower region of the validity of this treatment. 
We consider a background corresponding to a FLRW model filled with ultra-relativistic viscous QGP matter, whose bulk viscosity and equation of state have been deduced from recent heavy-ion collisions experiments and lattice QCD simulations. 

The present paper is organized as follows. The cosmological constant and the equation of state of viscous QGP are reviewed in section \ref{eosqgp}. Section \ref{field} is devoted to the evolution equations in viscous background QGP matter at finite cosmological constant. In section~\ref{approx}, we present analytical solutions for the evolution equations. The results and conclusions are given in section~\ref{final}.

\section{The cosmological constant and viscous QGP equation of state}
\label{eosqgp}

\subsection{The cosmological constant}

The cosmological constant $\Lambda$ was a subject of various modifications. Originally, it has been introduced to fit astrophysical data \cite{reef1}. The essential role of $\Lambda$ is to allow for a static solution to Einstein's equations, Eq. (\ref{ein}), in a cosmological background filled with matter has been fundamentally modified, especially at the time of the discovery of the universe expansion \cite{reef2}. 

The cosmic microwave background (CMB) observations indicate that the universe seems to have a negligible space curvature. Hereafter, we shall assume that the universe is spatially flat. Other astrophysical and cosmological data gathered over the last couple years apparently support the models with finite dark energy. An example can be taken from the redshift of supernovae \cite{sI1}. This apparently shows that the universe is currently undergoing an accelerated expansion. Furthermore, it seems that the flat universe contains both matter and cosmological constant $\Lambda$ so that $\Lambda=0.06^{+0.28}_{-0.34}$ or for a stringent upper limit $<0.51$ (at the $95\%$ confidence level). On the other hand, there are many observations indicating that the non-relativistic matter contributes with $\sim 1/3$ of the critical density \cite{refff1}. In other words, such observations would also mean that the universe contains $\sim 2/3$ of its total energy density in form of dark energy. The latter is a substance with negative effective pressure responsible for the {\it current} accelerated expansion \cite{refff2,refff3}. 

Although, the original motivation for the cosmological constant $\Lambda$ is not entirely clear \cite{refff5}, it may be connected with scalar fields \cite{refff6} (dilaton and moduli). Also, the interactions with the background strings may produce chaos \cite{refff7} and hence $\Lambda$ would gain small stochastic contributions \cite{refff8}. $\Lambda$ can be taken as a legitimate addition to the gravitational field equations, Eq. (\ref{ein}) for instance, or as a parameter to be constrained by observation, as mentioned before. The nature of dark energy itself is a subject of speculation. Apparently, it is very homogeneous but not very dense. It interacts through the gravitational force, exclusively. In order to determine the origin of $\Lambda$, all other terms in Eqs. (\ref{2}) and (\ref{3}) have to be measured to a sufficient precision. Such a way, one finds out that the $\Lambda/3$-term has to be finite. 

From theoretical point-of-view, $\Lambda$ can be interpreted as a measure of the energy density $\rho$ of the vacuum which is supported by the existence of {\it ''zero point''} energy in quantum mechanics based on particle-antiparticle pairs and confirmed through Casimir effect. The total energy density stems from various mostly unrelated sources. Each of them seems to come up with a contribution which is larger than the upper limits of today's $\Lambda$-value \cite{reef3a,reef4a}. In general relativity, all forms of $\rho$ are eligible to gravitate through modifying the space. $\Lambda$ is referring to $\rho^{vac}\equiv\Lambda/8\pi$ and a negative gravitation and therefore seems to explain the expansion nature of the {\it non-empty} universe. Furthermore, there is a recent suggestion that the velocity of light $c$ and the expansion of the universe are two aspects of one single concept connecting space and time in the expanding universe \cite{sI2}.

On the other hand, the role of cosmological constant in QGP era is conjectured to be played by the bag constant {\bf B} \cite{lambda1c}. It would be interesting to see what kind of long range effects (like interactions or correlations) {\bf B} may produce. For the high density QGP phase, it seems to be eligible to add a cosmological term with a large cosmological constant of the order of the bag constant $\bf B$ to the covariantly conserved energy-momentum tensor of a viscous fluid. At low temperatures ($<10\,$ MeV), a tiny cosmological constant is to be added to $T_{\mu\nu}$ of an ideal mixture. An understanding of the transition between these two descriptions is, of course, completely in the dark. This is actually an aspect of the profound $\Lambda$-problem. On the other hand, such a description would require to have equations of state describing the two phases (hadrons and QGP), simultaneously.  

Based on various experimental evidences, it is assumed that {\bf B} at very high temperatures likely disappears from the equation of state. At such high temperatures, the cosmological constant $\Lambda$ has to be introduced, explicitly. Also, it should be noticed that the structural latent heat of QGP would lead to small value of $\Lambda$. As will be given in Eqs. (\ref{2}) and (\ref{3}), $\Lambda$ is the only positive term in the field equations. Another  stunning fact is that $\Lambda$ remains constant, especially when all kinds of interactions between matter and radiation entirely disappear \cite{lambda1a}. Its role apparently depends on reconciling the age problem of the universe and the overall matter content \cite{lambda1b}. The latter can be dominant or minor relative to $\Lambda$ and therefore the evolution of different cosmological parameters would reflect the role of $\Lambda$. The presence of $\Lambda$ apparently influences the expansion of the universe, i.e. its age. 

The reasons why $\Lambda$ is to be included in QGP era are the long-range correlations and the constant self-interactions in QGP matter \cite{lambda1c}, besides the cosmological motivations. Recent results from RHIC approve the strong correlations in QGP phase \cite{reff1}.  The equation of state of dark energy can simply be characterized by negative $\omega$, Eq. (\ref{13}). In the case that $\omega=-1$, then 
\begin{eqnarray} \label{darkeq1}
t &=& \frac{\ln(a)}{H}.
\end{eqnarray}
Recent observations of supernova, galaxy clusters, and CMB seem to confirm that $\omega$ has values slightly deviate from $-1$ \cite{recnt}. In section that follows, the equation of state of viscous QGP is deduced.

\subsection{The equation of state of viscous QGP}

The equation of state, the temperature and the bulk viscosity of QGP, can be determined approximately at high temperatures~\cite{karsch07} from recent lattice QCD calculations~\cite{Cheng:2007jq}, as
\begin{eqnarray}
p &=& \omega \rho,\nonumber \\
T &=& \beta \rho^r, \label{13} \\
\xi &=& \alpha \rho^{\lambda}, \nonumber 
\end{eqnarray}
with $\omega = (\gamma-1)$, $\gamma \simeq 1.183$, $r\simeq 0.213$, $\beta\simeq 0.718$, 
$\alpha = (9\gamma^2-24\gamma+16)/((\gamma-1)\omega_0)$ 
and  $\omega_0 \simeq 0.5-1.5$ GeV. In order to close the system of the cosmological equations, we still need an expression for the relaxation time $\tau $ \cite{Ma95}, 
\begin{equation}\label{tau}
\tau=\frac{\xi}{\rho}=\alpha\,\rho^{\lambda-1}.
\end{equation}
Expressions (\ref{13}) and (\ref{tau}) are known as barotropic equations. Such a dependence has a practical advantage. As we give below, the field equations, Eq. (\ref{2}) and (\ref{3}), relate the cosmological parameters, like Hubble parameter $H$ and scale factor $a$, to the energy density $\rho$. Again, the expression (\ref{13}) are standard in analyzing the viscous cosmological models, whereas the equation for $\tau$ is a simple procedure to ensure that the speed of viscous pulses does not exceed the speed of light. 

Using this equation of state sets the validity of this treatment. It deals with viscous QGP matter. In present work, we mean with early universe a well-defined era, where the background matter is characterized by QGP.

\section{Evolution equations in viscous background QGP matter}\label{field}

We assume that the geometry of the early universe is flat and filled with a bulk viscous cosmological fluid, which can be described by a spatially flat FLRW type metric with $c=1$ given by
\begin{equation}  \label{1}
ds^{2}=dt^{2}-a^{2}(t)\left[dr^{2}+r^{2}\left(d\theta^{2}+\sin^{2}\theta d\phi^{2}\right)\right].
\end{equation}
The Einstein gravitational field equations in Robertson-Walker metric with a positive cosmological constant read
\begin{equation}
R_{ij}-\frac{1}{2}g_{ij}\,R+g_{ij}\Lambda=\frac{8\pi G}{c^4}\,T_{ij},  \label{ein}
\end{equation}
where $\Lambda$ is seen as a free parameter. 
The energy-momentum tensor of bulk viscous cosmological fluid filling the background geometry is given by
\begin{equation}
T_{i}^{j}=\left(\rho + p_{eff}\right) u_{i}u^{j}-p_{eff}\delta_{i}^{j}. \label{1_a}
\end{equation}
Following the Einstein's prescription that the cosmological constant can be treated as an independent parameter so that the energy-momentum tensor of background vacuum is given as 
\begin{equation} \label{cc1}
T_{ij}^{vac} = - \frac{g_{ij}}{8\pi}\, \Lambda,
\end{equation}
where $i$ and $j$ take $0,1,2,3$ and $\rho$ is the mass density. When taking Eq. (\ref{cc1}) into account,  Eq. (\ref{ein}) can be modified as follows. 
\begin{equation}
R_{ij}-\frac{1}{2}g_{ij}\,R=8\pi G\,\left(T_{ij}-g_{ij}\,\rho^{vac}\right),  \label{ein2}
\end{equation}
where $\rho^{vac}=\Lambda/8\pi$ is the vacuum energy density.

Generally, the bulk viscous effects can be described by means of an effective pressure $p_{eff}$ which includes the thermodynamic pressure $p$ and the bulk viscous pressure $\Pi$ \cite{Ma95}. The four velocity $u_{i}$ is conjectured to satisfy the condition $u_{i}u^{i}=1$. The particle and entropy fluxes characterize the background fluid, nearly entirely. Both are defined according to $N^{i}=nu^{i}$ and $S^{i}=sN^{i}-\left(\tau\Pi^{2}/2\xi T\right) u^{i}$, where $n$ is the number density, $s$ the specific entropy, $T\geq0$ the temperature, $\xi$ the bulk viscosity coefficient, and $\tau\geq0$ the relaxation coefficient for transient bulk viscous effect sometimes referred to as the relaxation time, Eq. (\ref{tau}). In ${\cal N}=4$ SYM theory, the relaxation time $\tau$ can be given by $(2-\ln 2)/(2\pi T)$ \cite{baier}. This has a very much little physical motivation. 

Taking into account a finite cosmological constant $\Lambda$, then in the comoving frame the energy momentum tensor has the components $T_{0}^{0}=\rho ,T_{1}^{1}=T_{2}^{2}=T_{3}^{3}=-p_{eff}$. For the line element given by Eq.~(\ref{1}), the Einstein field equations read
\begin{eqnarray}  \label{2}
3H^2 &=& 8\,\pi\,G\,\rho+\Lambda, \\
3\left(\dot{H}+H^2\right)&=&-4\,\pi\,G\,\left(\rho+3p_{eff}\right)+\Lambda, \label{3}
\end{eqnarray}
where an over-dot denotes the derivative with respect to the cosmic time $t$ and $G$ is the gravitational constant. The Hubble parameter $H$ is related to the scale factor $a$ via $H=\dot{a}/a$. In rest of this article, we take into consideration natural units, i.e. $c=\hbar=G=1$, for instance. The latter has been conventionally modified as, $(4\pi/3)G=1$. Therefore, in order to obtain the expressions given in Ref. \cite{Tawfik:2011bm}, $\Lambda$ and $\lambda$ should be assigned to $0$ and $1$, respectively. 

Assuming that the total matter content of the universe is conserved, $T_{i;j}^j=0$, then the evolution of the energy density of cosmic matter seems to fulfill the following conservation law:
\begin{equation}  \label{5}
\dot{\rho}=-3H\left(\rho+p_{eff}\right),
\end{equation}
which simply can be obtained from Eqs. (\ref{2}) and (\ref{3}). Apparently, this relation entirely eliminates the cosmological constant $\Lambda$.

\subsection{Eckart relativistic viscous fluid}

The first attempts at creating a theory of relativistic fluids were those of Eckart \cite{Ec40} and Landau and Lifshitz \cite{LaLi87}. These theories are now known to be pathological in several respects. Regardless of the choice of the equation of state, all equilibrium states in these theories are unstable and in addition signals may be propagated through the fluid at velocities exceeding the speed of light $c$. These problems arise due to the nature of the first order of this theory, that it considers only the first order deviations from the equilibrium leading to parabolic differential equations, hence to infinite speeds of propagation for heat flow and viscosity, in contradiction with the principle of causality. Conventional theory is thus applicable only to phenomena which are quasi-stationary, i.e. slowly varying on space- and time-scales characterized by mean free path and mean collision time.

The Eckart theory can be applied on modeling the cosmic background fluid as a continuum with a well-defined average 4-velocity field $u^{\alpha}$ where $u^{\alpha}u_{\alpha}=-1$. The vector number density $n^{\alpha}=nu^{\alpha}$ can be estimated, when unbalanced creation/annihilation processes take place; $n^{\alpha}_{;\alpha}=0$. This apparently means that  
\begin{equation}
 \dot n+3Hn=0,
\end{equation}
where the Hubble parameter $H=\nabla u$. In case of viscous fluid, the entropy current, 
\begin{eqnarray}
S^{\alpha}&=&snu^{\alpha},
\end{eqnarray}
is no longer conserved. The covariant form of second law of thermodynamics is $S^{\alpha}_{;\alpha}\ge 0$ and the divergence of entropy current is given by $T S^{\alpha}_{;\alpha}=-3H\Pi$. 

The evolution of the cosmological fluid is subject to the dynamical laws of
particle number conservation $N_{i}^{i}=0$ and Gibbs' equation
$Td\rho=d\left(\rho/n\right)+pd\left(1/n\right)  $.
In the following, we suppose that the energy-momentum tensor of the
cosmological fluid is conserved, $T_{i;l}^{l}=0$. Then, based on Gibbs' equation, the covariant entropy current can be achieved by  
\begin{equation} \label{entr}
\Pi=-3\xi H.
\end{equation}
This is a linear {\it first-order} relationship between the thermodynamical flux $\Pi$ and the corresponding force $H$. Substituting with the system of equations given in Eq. (\ref{13}) on the equation (\ref{2}) and (\ref{3}) results in   
\begin{eqnarray}
\dot H &=& 12 \pi \alpha \left(\frac{3 H^2 - \Lambda}{8 \pi}\right)^{\lambda}\, H -\frac{3}{2} \gamma H^2 + \frac{1}{2}\gamma \Lambda.
\label{Eck-Hh}
\end{eqnarray}
Analytical solutions for this differential equations will be introduced in section \ref{Eck-approx}.

\subsection{Israel-Stewart relativistic viscous fluid}
 
A relativistic second-order theory has been developed by Israel and Stewart \cite{Is76,IsSt76}, Hiscock and Lindblom \cite{HiLi89} through {\it ``extended''} irreversible thermodynamics. In this model, the deviations from equilibrium (bulk stress, heat flow and shear stress) are treated as independent dynamical variables, resulting in 14 dynamical fluid variables to be determined. The causal thermodynamics and its role in general relativity are reviewed in Ref. \cite{Ma95}. A general {\it algebraic} form for $S^{\alpha}$ including a {\it second-order} term in the dissipative thermodynamical flux $\Pi$ \cite{Is76,IsSt76} reads
\begin{equation}
S^{\alpha}=s n u^{\alpha}+\beta\Pi^2\frac{u^{\alpha}}{2T},
\end{equation}
where $\beta$ is a proportional constant.

For the evolution of the bulk viscous pressure, we adopt the causal evolution
equation \cite{Ma95} obtained in the simplest way (linear in $\Pi)$ to
satisfy the $H$-theorem (i.e., for the entropy production to be non-negative,
$S_{;i}^{i}=\Pi^{2}/\xi T\geq0$ \cite{Is76,IsSt76}). According to the causal relativistic IS theory, the evolution equation of the bulk viscous pressure reads~\cite{Ma95}
\begin{equation}  \label{8}
\tau \dot{\Pi}+\Pi =-3\xi H-\frac{1}{2}\tau \Pi \left( 3H+\frac{\dot{\tau}}{%
\tau }-\frac{\dot{\xi}}{\xi }-\frac{\dot{T}}{T}\right).
\end{equation}
In order to have a closed system from equations (\ref{2}) and (\ref{8}), we have to take into consideration equations of state for $p$ and $T$.

With the use of Eqs.~(\ref{8}), (\ref{13}) and (\ref{tau}), respectively, we obtain the following equation describing the cosmological evolution of the Hubble function $H$
\begin{eqnarray} \label{init}
\ddot{H} + \frac{3}{2\pi}\frac{B}{A} H \dot{H}^2 + 
\left(\frac{3(8\pi-B\gamma)}{2^{3\lambda}}+\frac{(B+6\pi\gamma)\Lambda}{A} - \frac{2\sqrt{C} A}{\alpha (-A)^{\lambda}} \frac{1}{H} - \frac{18\pi\gamma}{A}H^2\right) \frac{2^{3\lambda}}{16\pi}H \dot{H} &-& \nonumber \\
\frac{27 \gamma}{2^{4-3\lambda} \alpha} \frac{\sqrt{C}}{(-A)^{\lambda} A}\, H^6 - \frac{27(\gamma-2)}{4 A} H^5+\left(\frac{9}{4}\gamma-1\right)\frac{3\Lambda}{2^{2-3\lambda} \pi \alpha} \frac{\sqrt{C}}{(-A)^{\lambda} A}\, H^4 + 3\frac{\gamma B+4\pi(3\gamma-8)}{8\pi A} \Lambda H^3 &+&\nonumber \\
\left(1-\frac{9}{8}\gamma\right) \frac{\Lambda^2}{2^{1-3\lambda} \pi \alpha} \frac{\sqrt{C}}{(-A)^{\lambda} A}\, H^2+\frac{4 (B+15\pi)+3\gamma(B+6\pi)}{24\pi A} \Lambda^2 H &+& \nonumber \\
\left(\frac{\gamma}{8}-\frac{1}{6}\right) \frac{\Lambda^2}{2^{1-3\lambda} \pi \alpha} \frac{\sqrt{C}}{(-A)^{\lambda} A}\, &=& 0,
\end{eqnarray}
where $A=-H^2+\Lambda$, $B=1+r$ and $C=\pi^{2 \lambda}$. The analytical solutions are given in section \ref{IS-approx}.

\section{Analytical solutions at finite cosmological constant}\label{approx}

\subsection{Analytical solutions in Eckart relativistic viscous fluid} \label{Eck-approx}

At arbitrary values of the parameters $\Lambda$ and $\lambda$, the expression (\ref{Eck-Hh}) apparently turns to be non-integrable. Let us first check the default values $\Lambda=0$ and $\lambda=1$. The solutions simply read \cite{Tawfik:2011bm}
\begin{eqnarray} \label{eq:eckartIdeal}
t &=& \frac{3}{2\gamma^2}\, \frac{1}{H} \left[\gamma + 3 \alpha H \, \ln\left(3 \alpha - \frac{\gamma}{H} \right)\right].
\end{eqnarray}
On the other hand, when assuming that $\lambda=1+\varepsilon$, where $\varepsilon$ is a small positive value, then Eq. (\ref{Eck-Hh}) can be reduced to 
\begin{eqnarray}
\dot H &\approx& \frac{9}{2} \alpha\, H^3 - \frac{3}{2} \gamma\, H^2 - \frac{3}{2} \alpha \Lambda\, H + \frac{1}{2} \gamma \Lambda,
\end{eqnarray}
which can be solved as
\begin{eqnarray} \label{Ec:solC}
t &=& -\frac{1}{3 \gamma^2 - 9 \alpha^2 \Lambda} \left[ 2 \gamma \sqrt{\frac{3}{\Lambda}} \; \text{arctanh}\left(\sqrt{\frac{3}{\Lambda}}\,H \right) + 
   6\alpha \ln \left(\frac{3\alpha H-\gamma}{(3H^2-\Lambda)^{3\alpha}}\right) \right],
\end{eqnarray}
The graphical representations of Eqs. (\ref{eq:eckartIdeal}) and (\ref{Ec:solC}) are given in Fig. \ref{Figg3}. The first one seems to reflect the treatment of Ref. \cite{Tawfik:2011bm}, where the dark energy, $\Lambda$, has been excluded.

Although the mathematical difficulties arising with utilizing IS theory in the early universe, its application is simply unavoidable. It is derived by  the real concern about the causality constrains and non-consistency with the second law of thermodynamics.

\subsection{Analytical solutions in Israel-Stewart relativistic viscous fluid}
\label{IS-approx}

We introduce the transformation $u=\dot{H}$, so that Eq.~(\ref{init}) is transformed into a first order ordinary differential equation,
\begin{eqnarray}\label{init2}
u\frac{du}{dH} + \frac{3}{2\pi}\frac{B}{A} H u^2 + 
\left(\frac{3(8\pi-B\gamma)}{2^{3\lambda}}+\frac{(B+6\pi\gamma)\Lambda}{A} - \frac{2\sqrt{C} A}{\alpha (-A)^{\lambda}} \frac{1}{H} - \frac{18\pi\gamma}{A}H^2\right) \frac{2^{3\lambda}}{16\pi}H u &-& \nonumber \\
\frac{27 \gamma}{2^{4-3\lambda} \alpha} \frac{\sqrt{C}}{(-A)^{\lambda} A}\, H^6 - \frac{27(\gamma-2)}{4 A} H^5+\left(\frac{9}{4}\gamma-1\right)\frac{3\Lambda}{2^{2-3\lambda} \pi \alpha} \frac{\sqrt{C}}{(-A)^{\lambda} A}\, H^4 + 3\frac{\gamma B+4\pi(3\gamma-8)}{8\pi A} \Lambda H^3 &+&\nonumber \\
\left(1-\frac{9}{8}\gamma\right) \frac{\Lambda^2}{2^{1-3\lambda} \pi \alpha} \frac{\sqrt{C}}{(-A)^{\lambda} A}\, H^2+\frac{4 (B+15\pi)+3\gamma(B+6\pi)}{24\pi A} \Lambda^2 H &+& \nonumber \\
\left(\frac{\gamma}{8}-\frac{1}{6}\right) \frac{\Lambda^2}{2^{1-3\lambda} \pi \alpha} \frac{\sqrt{C}}{(-A)^{\lambda} A}\, &=& 0.
\end{eqnarray}
Using the new variable 
\begin{eqnarray}
\Omega &=& u \; E \; = u\; \exp\left(-\int \frac{3}{2\pi}\frac{B}{A} H dH\right),
\end{eqnarray}
we can rewrite Eq.~(\ref{init2}) in the form
\begin{equation} \label{OmegH1}
\Omega \frac{d\Omega }{dH} = F_1(H)\Omega + F_0(H),
\end{equation}
where
\begin{eqnarray}
F_1(H) &=& \frac{2^{-4+3\lambda} (-A)^{\frac{B}{4\pi}}}{\pi} \left[3\, 8^{-\lambda} (8\pi-B\gamma)+\frac{(B+6\pi\gamma)\Lambda}{A}-\frac{2 A \sqrt{C}}{\alpha (-A)^{\lambda} H}-\frac{18 H^2 \pi \gamma}{A}\right]\, H, \\
F_0(H) &=& -\frac{(-A)^{-1+\frac{B}{2\pi}-\lambda}}{48\pi \alpha} \left[8^{\lambda}\sqrt{C}\left(-81 H^6 \pi \gamma+9 H^4 (-4+9\gamma)\Lambda +\left(-4+3 H^2 (8-9\gamma)+3\gamma \right) \Lambda^2\right) \right. \nonumber \\ 
 & & \left.  - 2(-A)^{\lambda} H \alpha \left(162 H^4 \pi (-2+\gamma)-9 H^2 (B\gamma+4 \pi(-8+3\gamma)) \Lambda-(4B+60\pi+3(B+6\pi)\gamma) \Lambda^2\right)\right].
\end{eqnarray}
By introducing a new independent variable $z=\int F_1(H)\,dH$, we obtain
\begin{equation}
\Omega \frac{d\Omega}{dz} - \Omega = g(z),
\end{equation}
where $g(z)=F_0/F_1$ can be approximated as a simple function depending on the new variable $z$ as follows.

\begin{figure}
\includegraphics[width=8cm,angle=0]{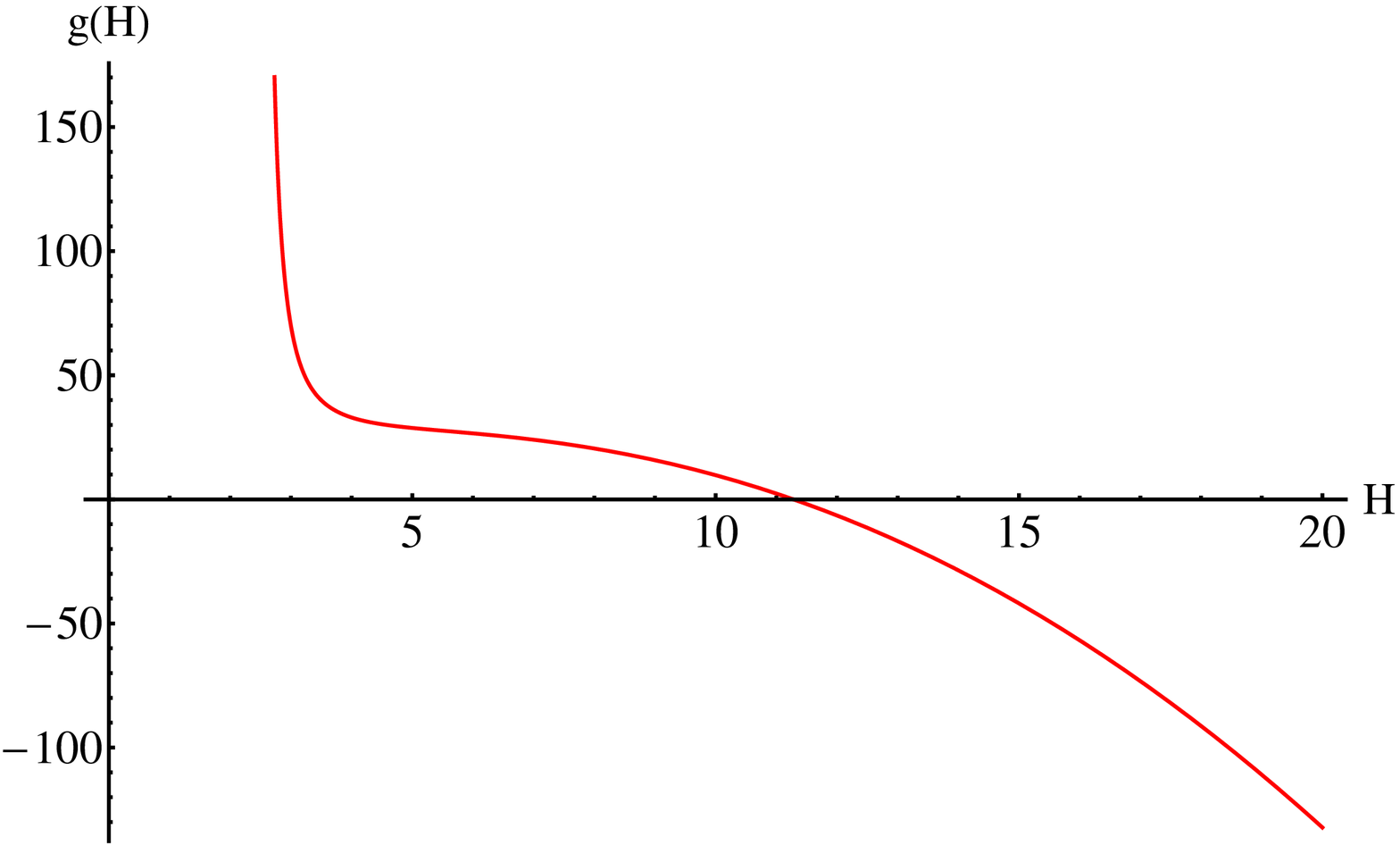}
\includegraphics[width=8cm,angle=0]{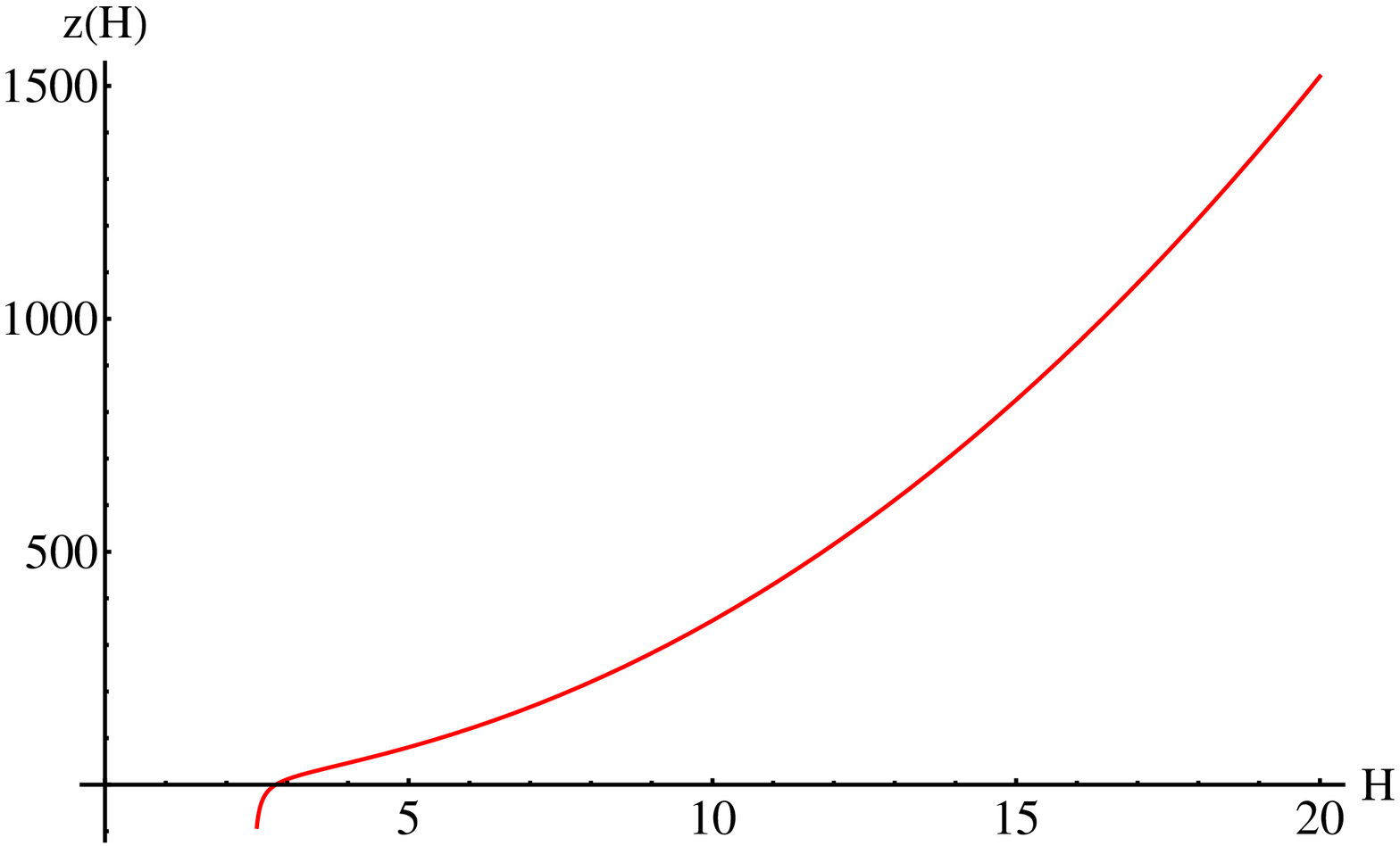}
\caption{In left panel, $g(H)=F_0/F_1$ is given in dependence on $H$, whereas in right panel $z(H)$ is depicted. }
\label{Figg1}
\end{figure}

\begin{figure}
\includegraphics[width=14cm,angle=0]{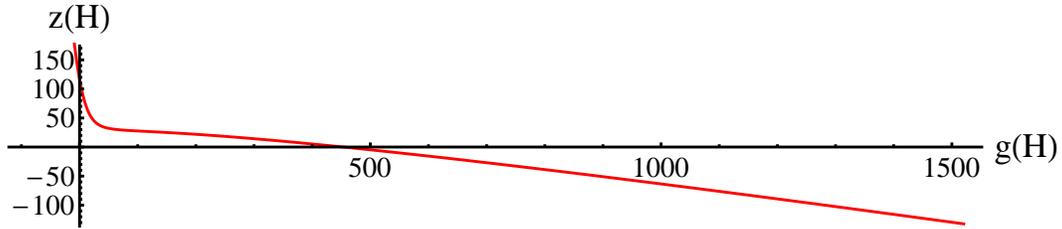}
\caption{Parametric plot of $g(H)$ versus $z(H)$. The function $g$ seems to depend on $z$, linearly.}
\label{Figg2}
\end{figure}

\begin{eqnarray} \label{fullgofz}
g(H)&=& (-A)^{\frac{B}{4\pi}} 
\frac{8^q \sqrt{C} \left(-81 H^6 \pi \gamma +9 H^4 (9\gamma-4)\Lambda + \left(3 H^2 (8-9 \gamma)+3\gamma - 4\right) \Lambda^2\right)}{-3\, 2^{1+3 q} A^2 \sqrt{C}+3 (-A)^q H \alpha  \left(3 A (8 \pi -B \gamma )+8^q \left(B \Lambda +6 \pi  \gamma  A \right)\right)} - \nonumber \\
& & (-A)^{\frac{B}{4\pi}} \frac{2 (-A)^q H \alpha \left(162 H^4 \pi (\gamma-2) - 9 H^2 (B \gamma + 4\pi(3\gamma-8)) \Lambda - (4B+60\pi + 3(B+6\pi)\gamma)\Lambda^2\right)}{-3\, 2^{1+3q} A^2 \sqrt{C}+3 (-A)^q H \alpha  \left(3 A (8\pi-B\gamma) + 8^q \left(B \Lambda + 6\pi \gamma A \right)\right)}, \\
z(H) &=& \frac{2^{3\lambda-5}}{\pi} \frac{(-A)^{\frac{B}{4\pi}}}{A} \left[-9 \pi \gamma H^3 + \left(\frac{3}{8^{\lambda}} A (8 \pi-B\gamma) + (B+6\pi \gamma) \Lambda \right) H -\frac{4}{\alpha}\frac{\sqrt{C}}{(-A)^{\lambda-2}} \right]\, H. \label{Eq2}
\end{eqnarray}
Apparently, replacing $H$ with $z$ is a non-trivial task. To this goal, let us first depict the real parts of both functions $g(H)$ and $z(H)$, Fig. \ref{Figg1}. Obviously, the two functions have various ingredients. Now, we may try the  parametric plotting, as in Fig. \ref{Figg2}. It is obvious that the dependence of $g$ on $z$ can approximately be characterized as linear. 
\begin{equation}
g(z)\approx \phi \, z,
\end{equation}
where $\phi=-0.07\pm0.02$. Then, from the definition of $\Omega$, we simply have
\begin{eqnarray}
\Omega &=& \left(3 H^2-\Lambda\right)^{-B/4\pi}\, \dot{H}. \label{Eq1} 
\end{eqnarray}

Now, we reduce the last expression to the canonical equation of the Abel type. In doing this, we use the relation $\Omega = z/{\cal P}$. Then from Eqs.~(\ref{Eq1}) and (\ref{Eq2}), we obtain a first order differential equation for $H$. 
\begin{eqnarray} \label{init-polyn}
{\cal P} \dot H &=& \frac{2^{3\lambda-5}}{\pi} \, H \left[- 9 \pi \gamma H^3 + \left(\frac{3}{8^{\lambda}} A (8 \pi - B \gamma) + \Lambda (B+6\pi\gamma)\right) H -\frac{4\sqrt{C}}{\alpha} (-A)^{2-\lambda}\right].
\end{eqnarray}
This equation has the solution
\begin{equation}
t = \frac{{\cal P}}{\alpha\, \Lambda\, {\cal N}}\, \left\{\left[3\alpha (B\gamma-8\pi)+8^{\lambda} \left(12\sqrt{C}(\lambda-2) \ln(H)\, H - \alpha (B+6\pi\gamma)\right)\right] \frac{32\pi}{H} - 348 \pi\right\}, \label{eq:mysolut1}
\end{equation}
where ${\cal N}=\left(24\pi-3B\gamma + 8^{\lambda} (B+6\pi\gamma)\right)^2$ and ${\cal P}$ is taken as a free parameter. We can assign any real value to ${\cal P}$. For the results presented in this work, we used a negative value. This negative sign is necessarily to overcome the sign from the integral limits.

\section{Results and Conclusions}\label{final}

In the present work, we have considered the evolution of a full causal bulk viscous flat, isotropic and homogeneous universe with finite bulk viscosity parameter ($\xi\neq 0$) and and cosmological constant ($\Lambda\neq 0$). The equation of state of QGP is taken from recent lattice QCD simulations and heavy-ion collisions.

\begin{figure}
\includegraphics[width=14cm,angle=0]{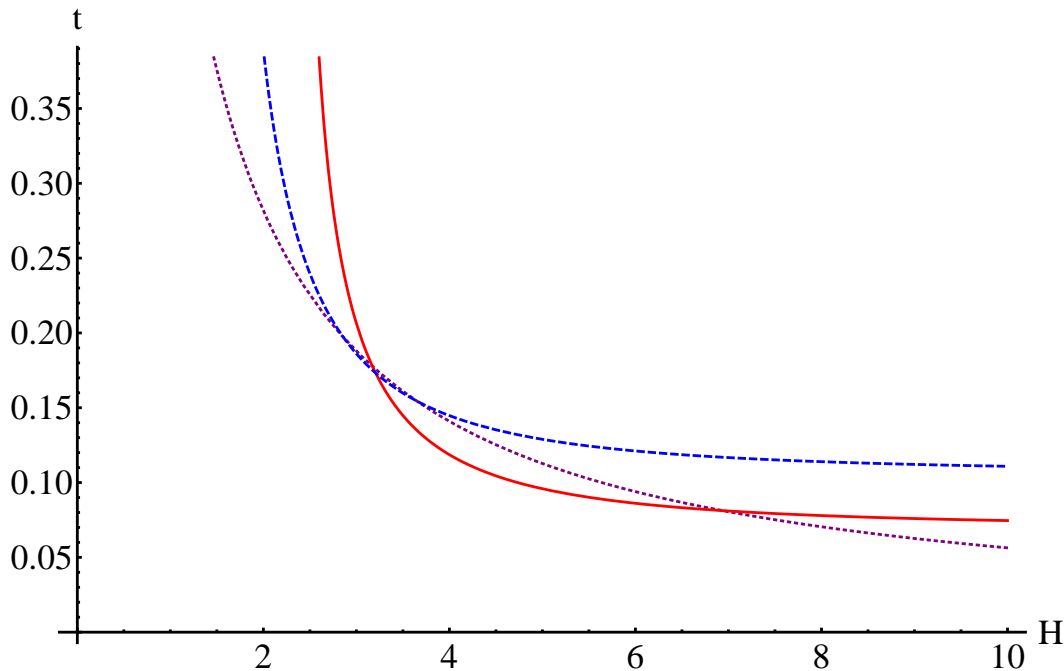}
\caption{Cosmic time $t$ vs. Hubble parameter $H$ in cosmic background filled with Eckart fluid. The solid curve represents the solution in Eq. (\ref{Ec:solC}). The solution at vanishing viscous and $\Lambda$ is given by the dotted curve. The dashed curve represents Eq. (\ref{eq:eckartIdeal}), where $\Lambda=0$ and $\lambda=1$, i.e. at vanishing dark energy.}
\label{Figg3}
\end{figure}

In Fig. \ref{Figg3}, the cosmic time $t$ and Hubble parameter $H$ in cosmic background filled with the Eckart fluid are depicted. In this treatment, $t$ is given in GeV$^{-1}$ whereas $H$ in GeV. The solid curve represents the analytical solution, Eq. (\ref{Ec:solC}). Using equation of state of an ideal and non-viscous gas and a vanishing cosmological constant $\Lambda$ result in the dotted curve. The dashed curve gives that results when $\Lambda$ is vanishing and $q=1$. This has been introduced in Ref. \cite{Tawfik:2011bm}. In all curves, the singularities are present. On the other hand, the time evolution shows essential differences. Eckart fluid enables $H$ to decay much faster (solid and dashed curves), especially at small $t$. This behavior is flipped at large $t$-values. This might reflect the effects of the approximation in $\lambda=1+\epsilon$, where $\epsilon$ is taken to be very small. 

\begin{figure}
\includegraphics[width=14cm,angle=0]{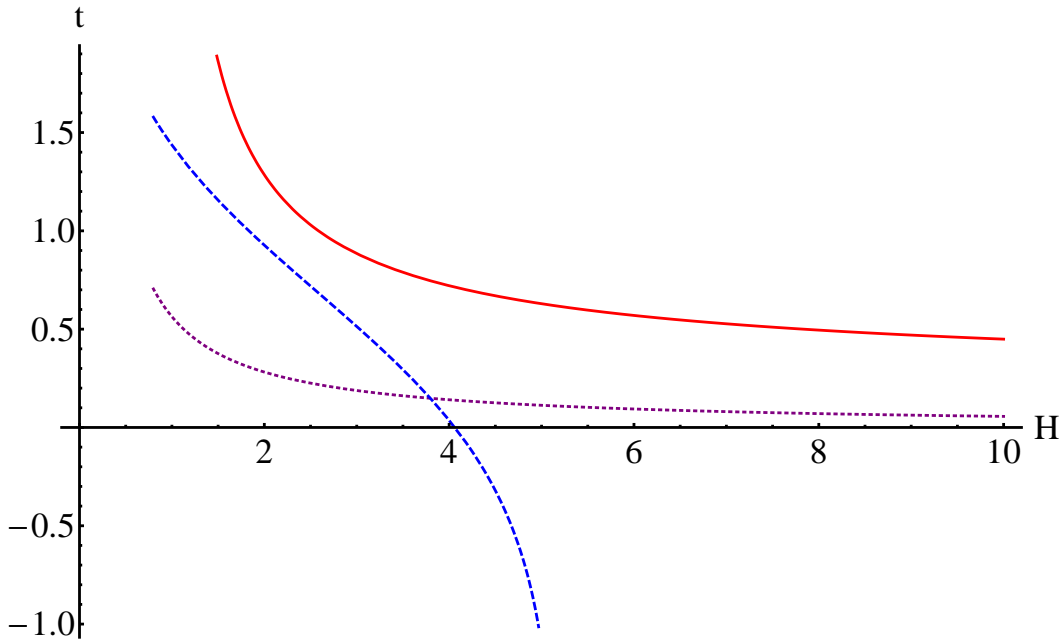}
\caption{Cosmic time $t$ vs. Hubble parameter $H$ in cosmic background filled with IS fluid. The solid curve represents the solution in Eq. (\ref{eq:mysolut1}). The dashed curve depicts the results of Ref. \cite{Tawfik:2011bm}, where $\xi\neq 0$ but $\Lambda=0$. The solution at vanishing viscous and $\Lambda$ is given by the dotted curve. Obviously, the negative values are non-physical.}
\label{Figg4}
\end{figure}

In Fig. \ref{Figg4}, the analytical solution given in Eq. (\ref{eq:mysolut1}) is graphically illustrated. To have a comparison with other cases, we plot both non-viscous and viscous solutions. The dashed curve gives the latter case, where the cosmological constant is vanishing \cite{Tawfik:2011bm}. 
\begin{eqnarray} \label{eq:nonviscousAA}
t &=&  \ln \left[ \frac{1}{\alpha r\, H} - \frac{3}{2} \left( \frac{1}{1-r} + \gamma \right) \right]^{- \alpha\, r\, {\cal P}}.
\end{eqnarray}
It is obvious that the causal bulk viscous universe described by this solution starts its evolution from an initial non-singular state with a non-zero initial value of $H(t)$. The ideal case is characterized by vanishing viscosity coefficient $\xi$ and cosmological constant $\Lambda$,
\begin{eqnarray}
t &=&  \frac{2}{3\, \gamma\, H}.
\end{eqnarray}
The singularity is apparently present.

So far we conclude that the bulk viscosity seems to play an important role in the evolution of the early universe. Despite of the simplicity of our model, it shows that a better understanding of the dynamics of our universe is only accessible, if we use reliable equation of state in order to characterize the matter filling the cosmic background geometry. Also, the cosmological constant seems to be essential. The motivation of finite $\Lambda$ is based on theoretical needs and astrophysical and cosmological observations \cite{reef1,reef2,sI1,refff1,refff2,refff3,sI2}. It is very essential to describe the early stages of universe, Eq. (\ref{darkeq1}).  

Assuming that $\Lambda$ vanishes (Fig. \ref{Figg4} and Eq. (\ref{eq:nonviscousAA})), the solution shows an absence of singularity near $t=0$. This was the main result of Ref. \cite{Tawfik:2011bm}. Whether it has a physical meaning, we like to refer to the restriction to the QCD era of early universe. It should not be understood that it is applicable either in former or later eras. By way of precaution, the QCD era can be limited to the temperatures in the range $0.2\leq T\leq10\;$GeV or time in the range $18.35\leq t \leq 0.0073\;$GeV$^{-1}$. Apparently, this limitation is likely compatible with the recent lattice QCD simulations and the experimental evidences that QGP seems to remain strongly correlated, i.e. non-viscous, up to several $T_c$.

In the present work, it is assumed that the universe is spatially flat and the background geometry is filled with dark energy and QCD matter (QGP) with a finite viscosity coefficient. The results seem to support the singularity near $t\rightarrow 0$. The resulting universe is obviously characterized by almost the same behavior of the ideal case. There is two essential differences. First, the solution given in Eq. (\ref{eq:mysolut1}) depends on free parameter ${\cal P}$. The solid curve can be moved up and down depending on ${\cal P}$. Its shape is not depending on it. Second, the universe in which dark energy is taken into account seems to live much longer that the ideal case.     

The validity of our treatment depends on the validity of the equations of states, Eq.~\ref{13}, which we have deduced from the lattice QCD simulations at temperatures larger than $T_c\approx 0.19~$GeV. Below $T_c$, as the universe  cooled down, not only the degrees of freedom suddenly increase~\cite{Tawfik03} but also the equations of state turn out to be the ones characterizing the hadronic matter. Such a phase transition - from QGP to hadronic matter - would characterize one end of the validity of our treatment. The other limitation is the very high temperature (energy), at which the strong coupling $\alpha_s$ entirely vanishes.


\end{document}